\documentclass[12pt]{article}
\usepackage{graphics}
\usepackage{epsfig}
\usepackage{amsmath}
\begin{document}
\begin{flushright}
revised 15 September, 2000\\
accepted for publication in {\em Physical Review D}\\
\end{flushright}
\vspace{.5in}
\begin{center}
{\Huge Graviton Production in Relativistic\\Heavy-Ion Collisions}\\
\vspace{.25in}
{\Large Sean C. Ahern, John W. Norbury and William J. Poyser}\\
\vspace{.15in}
{\em Physics Department, University of Wisconsin - Milwaukee,}\\
{\em P.O. Box 413, Milwaukee, Wisconsin 53201, USA.}\\
(email: sahern@uwm.edu; norbury@uwm.edu; poyser@uwm.edu)\\
\vspace{.25in}
PACS numbers: 25.75.-q, 25.20.Lj, 04.50.+h\\
\end{center}
\vspace{.75in}

\begin{center}
ABSTRACT
\end{center}
\medskip
\noindent
We study the feasibility of producing the graviton of the novel Kaluza-Klein
theory in which there are $d$ large compact dimensions in addition to the
$4$ dimensions of Minkowski spacetime. We calculate the cross section for
producing such a graviton in nucleus-nucleus collisions via t-channel
$\gamma \gamma$ fusion using the semiclassical Weizs\"{a}cker-Williams
method and show that it can exceed the cross section for graviton production
in electron-positron scattering by several orders of magnitude. 

\newpage
\newcounter{chapter}
\setcounter{chapter}{1}
\begin{center}
\Roman{chapter}. INTRODUCTION
\end{center}
\stepcounter{chapter}
\medskip
Until recently, it has been believed that the role of gravity in particle
interactions only becomes important at the experimentally inaccessible energy
scale of the Planck mass, $M_P=1.2\times10^{19} \ GeV$. Recent advances in
M-Theory, a Kaluza-Klein (KK) theory in which there are 11 spacetime dimensions,
suggest that there might be an effective Planck mass $M_D$ much
lower than $M_P$ (perhaps as low as $O(1 \ TeV)$) at which
gravitational effects become important (cf. \cite{Ark,Han,Atw} and references
therein). Traditional KK theories contain the usual four dimensions of space
and time, plus additional compact dimensions which form an unobservably
small (perhaps Planck length sized) manifold. These recent models propose that
there are $d$ extra compact dimensions, all of roughly the same size $R$, but
$R$ might possibly be much larger than the Planck length --- perhaps as large
as a millimeter. The size $R$ of these $d$ dimensions is related to $M_P$
and $M_D$ via
\begin{equation}
R^d \sim \frac{M_P^2}{M_D^{d+2}}.
\label{one}
\end{equation}
$R$ also represents the scale at which the Newtonian inverse-square force law
is expected to fail. If $d=1$ and $M_D$ is taken to be $1 \ TeV$ then
$R\sim10^{10} \ km$, which implies there should be deviations from Newtonian
gravity over solar system distances. This case is clearly ruled out, but if $d\ge2$
then $R<1 \ mm$ --- a possibility that is not in conflict with any current
experimental data \cite{Ark,Atw}. While gravity has not been tested at distances
smaller than a millimeter, Standard Model (SM) interactions have been accurately
tested down to distances of about $10^{-16} \ cm$ \cite{Pol}. Hence the $4+d$
dimensional graviton is conjectured to propagate in the $4+d$ dimensional
spacetime bulk, while the SM particles are confined to a $4$ dimensional
submanifold (Minkowski spacetime).

\begin{figure}[t]
\begin{center}
\epsfig{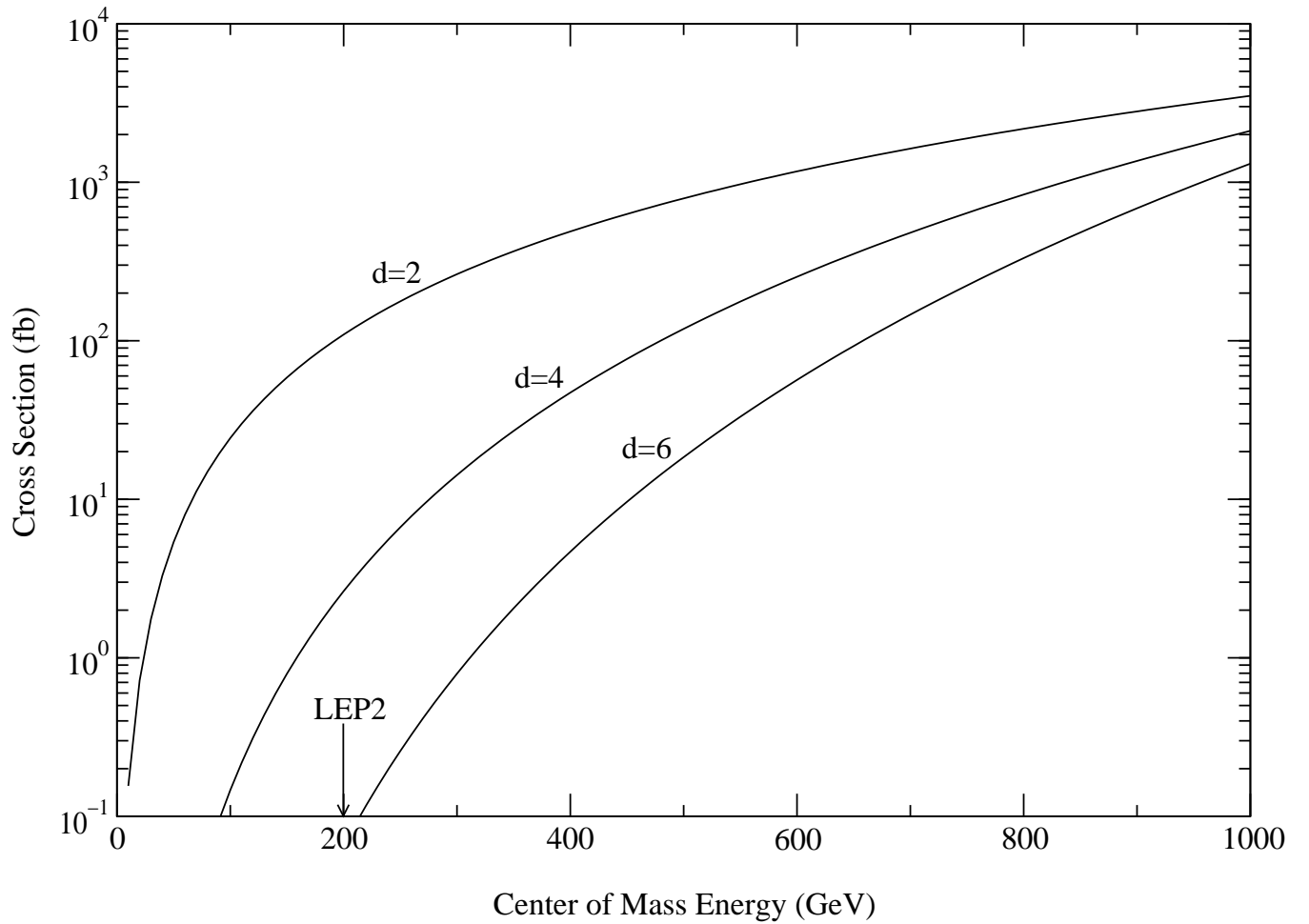}
\end{center}
\renewcommand{\figurename}{FIG.}
\caption{\label{eeG} Cross section for graviton production in $e^+e^-$ scattering ($e^+e^- \to e^+e^- \gamma \gamma \to e^+e^-G$) at three different values of $d$, the number of large compact dimensions.}
\end{figure}

Recently it has been shown (cf. \cite{Atw}) that such a graviton might be
detected as missing energy in electron-positron scattering. We propose here a
new way that the graviton might be discovered, namely in a peripheral (near-miss)
collision of two heavy nuclei. In Section II the basics of this novel KK scenario
are summarized.
In Section III, the calculation of the cross section for graviton
production in electron-positron scattering is summarized using the Feynman rules
in the Weizs\"{a}cker-Williams leading log approximation, as outlined in
\cite{Atw}.
Then, in Section IV, the process is generalized in the semiclassical
Weizs\"{a}cker-Williams method to graviton production via nucleus-nucleus
collisions.
\medskip
\begin{center}
\Roman{chapter}. SUMMARY OF THE NEW KALUZA-KLEIN SCENARIO
\end{center}
\stepcounter{chapter}
\medskip
The starting point of the novel KK scheme is a $4+d$ dimensional spacetime
action that describes the $4+d$ dimensional graviton fields. The action is
extremized and the theory is expanded around a vacuum metric that is the
product of Minkowski spacetime with a $d$ dimensional torus. The dependence
on the compact dimensions $x^d$ of the $4+d$ dimensional graviton is Fourier
expanded in a complete set of plane waves. Due to the topology of the torus, all
$d$ compact dimensions are periodic and consequently the wave numbers $k_n$
of the modes on the torus are all quantized: $k_n=2\pi n/R$, where $n$ is an
integer that labels the modes and the sizes of the $d$ compact dimensions are
assumed to all be $\sim R$. The $n=0$ modes, which are identically the
coefficients of the normal modes, are confined to Minkowski spacetime and
naturally divide into a massless spin-2 graviton (that gives rise to Newtonian
gravity), $d$ massless $U(1)$ gauge bosons and $d(d+1)/2$ massless scalar
bosons. The $n\ne0$ modes reorganize themselves at each KK level $n$ into a
massive spin-2 boson, $(d-1)$ massive vector bosons and $d(d-1)/2$ massive
scalar bosons, all of which have the same mass-squared, $m_n^2=4\pi^2n^2/R^2$.
The reorganization of these modes is associated with spontaneous
symmetry breaking where, like in the Higgs mechanism, the massless spin-2
graviton fields absorb the spin-1 and spin-0 fields and become massive.
This KK formalism is to be regarded as an effective theory, and an
ultraviolet cutoff $\Lambda \sim M_D$ is imposed on the tower of KK modes, so
that $m_n^2 < \Lambda^2$ for all $n$ \cite{Ark,Han}. In practical
calculations, such as the one in this study, a density $\rho$ of modes
(i.e., the differential number $dN$ per unit mass-squared $dm_n^2$) is
used. This function is derived in \cite{Han} and is given by	
\begin{equation}
\rho (m_n^2) \equiv \frac{dN}{dm_n^2 } = \frac{R^d m_n^{d-2}}{(4\pi)^{d/2}\Gamma (d/2)},
\label{two}
\end{equation}
where $\Gamma$ is the Gamma function. As pointed out in \cite{Han}, it is
this function that is to be convoluted with a physical amplitude or cross
section for a mode with mass $m_n$. While the coupling of any individual
mode to SM matter is Planck-mass suppressed (viz., $\propto M_P^{-2}$), the
``end-of-the-day'' coupling is only $M_D$ suppressed (viz.,
$\propto M_D^{-(d+2)}$) due to the factor of $M_P^2$ in $\rho$ (plug
(\ref{one}) into (\ref{two})) that multiplies each individual mode coupling
term in the overall sum. This enhancement is simply the result of the high
multiplicity of graviton states, as described by $\rho$, within the same
mass interval.
\medskip
\begin{center}
\Roman{chapter}. GRAVITON PRODUCTION IN ELECTRON-POSITRON SCATTERING
\end{center}
\stepcounter{chapter}
\medskip
The process considered in this section is graviton G production in electron-positron
scattering via t-channel $\gamma \gamma$ fusion (i.e., $e^+e^- \to e^+e^-
\gamma \gamma \to e^+e^-G$), which, according to \cite{Atw}, is a promising mechanism by which the particle might be produced and detected.
The t-channel process is more significant than the s-channel one, wherein the initial electron and positron mutually
annihilate, because the photons are produced almost collinearly with the electrons,
and are more likely to directly interact with one another than are the electron
and positron with each other. One way to determine the cross section for this
process is to use the Feynman rules with the Weizs\"{a}cker-Williams leading
log approximation. The calculation is performed in \cite{Atw} and reduced
to the following one dimensional integral:
\begin{equation}
\sigma_{\gamma\gamma} (e^+e^- \to e^+e^-G) =\frac{\alpha^2}{8s}\frac{\pi^{d/2-1}}{\Gamma (d/2)}\left[{\frac{\sqrt s}{M_D}} \right]^{d + 2}F_{d/2} \log ^2 \left[{\frac{s}{4m_e^2 }}\right].
\label{three}
\end{equation}
Here $\alpha \approx 1/137$ is the fine structure constant, $\sqrt s$ is the
center-of-mass energy of the collision and
$F_k\equiv \int_0^1 d\omega {f(\omega )\omega ^k}$, where
$f(\omega)=-\left[{(2+\omega)^2\log (\omega)+2(1-\omega)(3+\omega)}\right]/\omega$.
$\sigma_{\gamma\gamma}(e^+e^- \to e^+e^-G)$ is a function of input parameters
$\sqrt s$ and $d$. A plot of this function vs. $\sqrt s$ at three different
values of $d$ ($2$, $4$ and $6$) is shown in Fig. \ref{eeG}, along with an
indication of the operational energy of LEP2 (200 GeV).
\medskip
\begin{center}
\Roman{chapter}. GRAVITON PRODUCTION IN NUCLEUS-NUCLEUS COLLISIONS
\end{center}
\stepcounter{chapter}
\medskip
We now consider the possibility of producing a graviton in the
peripheral collision of two relativistic heavy ions, via t-channel
$\gamma\gamma$ fusion (i.e., $A_1A_2 \to A_1A_2 \gamma\gamma \to A_1A_2G$).
The colliding nuclei are taken to be identical and are described by the atomic number $Z$ (the number of protons) and the mass number $A$ (the number of protons and neutrons).
Our interest is motivated by the fact that cross sections for such reactions scale as $Z^4$, where (for heavy ions) $Z$ can be on the order of 10--100.
The Feynman rules are of course an extremely accurate way of determining cross
sections for simple processes, but are impractical for processes such as
nucleus-nucleus collisions, where many particles are interacting with
each other simultaneously. For these types of interactions, certain
approximations must be made in order to simplify the calculations.
For this study, we use one such approximation scheme --- the semiclassical
Weizs\"{a}cker-Williams method, wherein it is assumed that the colliding particles are ultrarelativistic and follow straight-line classical trajectories, and the photons mediating the interactions are real (``on-shell'').
In using this approach, the $\gamma\gamma \to G_n$ subprocess cross section for one KK mode is folded with the two separate Weizs\"{a}cker-Williams photon spectra $N(\omega_1)$ and $N(\omega_2)$, and the resulting quantity is summed over all contributing modes.
The $\gamma\gamma \to G_n$ subprocess cross section, denoted $\sigma_{\gamma_1\gamma_2 \to G_n}(\omega_1,\omega_2)$, is the cross section
for the production of one mode of mass $m_n$ via the fusion of two photons
of angular frequencies $\omega_1$ and $\omega_2$.
This function, which is easily derived from the Feynman rules, is found to be \cite{Bar}:
\begin{equation}
\sigma_{\gamma_1 \gamma_2 \to G_n}(\omega_1,\omega_2) = \frac{10\pi^2 }{m_n^2 }\Gamma_{G_n  \to \gamma \gamma} \delta (\sqrt{\hat s}-m_n).
\label{four}
\end{equation}
$\Gamma_{G_n \to \gamma\gamma}$ is the partial decay width for one graviton
mode to decay into two photons, $\delta$ is the Dirac delta function and
$\sqrt{\hat s}$ is the center of mass energy of the two-photon system.
$\Gamma_{G_n \to \gamma\gamma}$ is given in \cite{Han}, as:
\begin{equation}
\Gamma _{G_n  \to \gamma \gamma} = \frac{1}{20}\frac{m_n^3}{M_P^2}.
\label{five}
\end{equation}
The function $N(\omega)$ gives the number of virtual photons per unit photon
frequency outside the nucleus and can be derived from the Feynman rules or via
a classical analysis \cite{Ter,Kra,Vid,Jac,Eic}. Here we use a classical
formula, derived for example in \cite{Jac} and \cite{Eic}:
\begin{equation}
N(\omega) = \frac{2}{\pi}\frac{Z^2\alpha}{\omega \beta^2}\left\{{\xi K_0(\xi)K_1(\xi) - \frac{1}{2}\beta^2 \xi^2 \left[{K_1^2(\xi)-K_0^2(\xi)} \right]} \right\}.
\label{six}
\end{equation}
The functions $K_0$ and $K_1$ are modified Bessel functions of the second kind,
of order zero and one, respectively. The argument $\xi$ of these functions is
defined as $\xi \equiv \frac{\omega b_{\min}}{\gamma \beta}$, where $\beta$ is
the speed of either ion, $\gamma \equiv \frac{1}{\sqrt{1-\beta^2}}$ and 
$b_{\min}$ is the minimum impact parameter of the collision, which is the
distance of closest approach between the center of either nucleus and the point of G production.
We take $b_{\min}$ to be the nuclear radius so as to trigger against the strong interaction effects which completely dominate electromagnetic effects when the nuclei overlap \cite{Nor}.
For a nucleus of mass number $A$, the nuclear radius is $r \approx 1.2A^{1/3}\ fm$ \cite{Bau}. For our applications, we consider lead ($Z=82$ and $A=208$) and calcium ($Z=20$ and $A=40$) nuclei; thus
$b_{\min} \approx 7.11\ fm$ for the former and $b_{\min} \approx 4.10\ fm$
for the latter. Lead is interesting because it is one of the more common
stable nuclei with a high $Z$ value, and calcium is interesting because
it has a relatively high luminosity when used as a heavy ion beam \cite{Bra}.
The total cross section for the production of one mode of mass $m_n$ is: 
\begin{equation}
\sigma_{\gamma \gamma} (A_1A_2 \to A_1A_2G_n) = \int {d\omega_1 \int {d\omega_2 \ N(\omega_1) N(\omega_2)}} \sigma_{\gamma_1 \gamma_2 \to G_n}(\omega_1,\omega_2).
\label{seven}
\end{equation}
The limits of integration can be derived from conservation of 4-momentum;
an elegant version of such a calculation is presented in \cite{Cah}.
As used here, $m_n^2/2 \sqrt s \le \omega_1 \le \sqrt s/2$ and
$m_n^2/4 \omega_1 \le \omega_2 \le \sqrt s/2+m_n^2/2 \sqrt s-\omega_1$, but
because of the delta function in (\ref{four}), only the limits on
$\omega_1$ are needed. Finally, the total cross section for graviton
production in this process is found by summing (\ref{seven}) over all
contributing modes:
\begin{equation}
\sigma_{\gamma \gamma}(A_1A_2 \to A_1A_2G) = \int {dm_n^2 \ \rho(m_n^2)} \sigma_{\gamma \gamma}(A_1A_2 \to A_1A_2G_n).
\label{eight}
\end{equation}
$\rho(m_n^2)$ was given in (\ref{two}) and $m_n^2$ ranges from 0 to the
smaller of $s$, as demanded by conservation of energy and momentum, and
$\Lambda^2$, which is the absolute upper limit on $m_n^2$. In its simplest
form, the cross section is given by:
\begin{equation}
\sigma_{\gamma \gamma}(A_1A_2 \to A_1A_2G) = const. \int {dm_n \int
{d\omega_1 \ \frac{m_n^{d-1}}{\omega_1}f(\omega_1)f \left({\frac{m_n^2}{4\omega_1}} \right)}},
\label{nine}
\end{equation}
where
\begin{equation}
const. \equiv \frac{8}{(4 \pi)^{d/2} \Gamma(d/2)}\left({\frac{Z^2 \alpha}
{\beta^2}} \right)^2 \frac{1}{M_D^{d+2}}
\label{ten}
\end{equation}
and
\begin{equation}
f(\xi) \equiv \xi K_0(\xi)K_1(\xi) - \frac{1}{2}\beta^2 \xi^2
\left[{K_1^2(\xi)-K_0^2(\xi)} \right]
\label{eleven},
\end{equation}
and the limits of integration and values of parameters are as specified above.

\begin{figure}[t]
\begin{center}
\epsfig{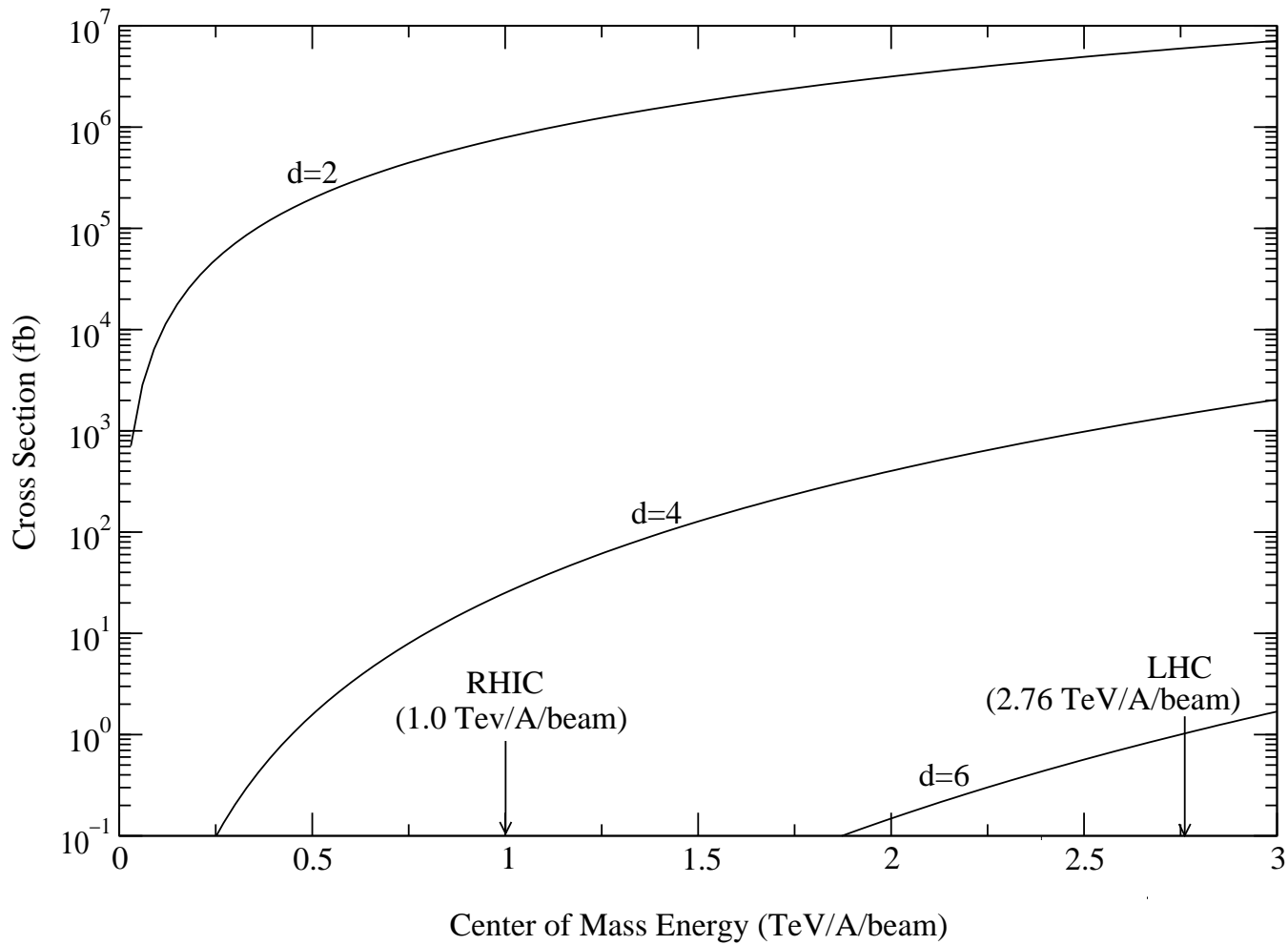}
\end{center}
\renewcommand{\figurename}{FIG.}
\caption{\label{pbpbG} Cross section for graviton production in Pb-Pb collisions ($PbPb \to PbPb \gamma \gamma \to PbPbG$) at three different values of $d$, the number of large compact dimensions.}
\end{figure}

\begin{figure}[t]
\begin{center}
\epsfig{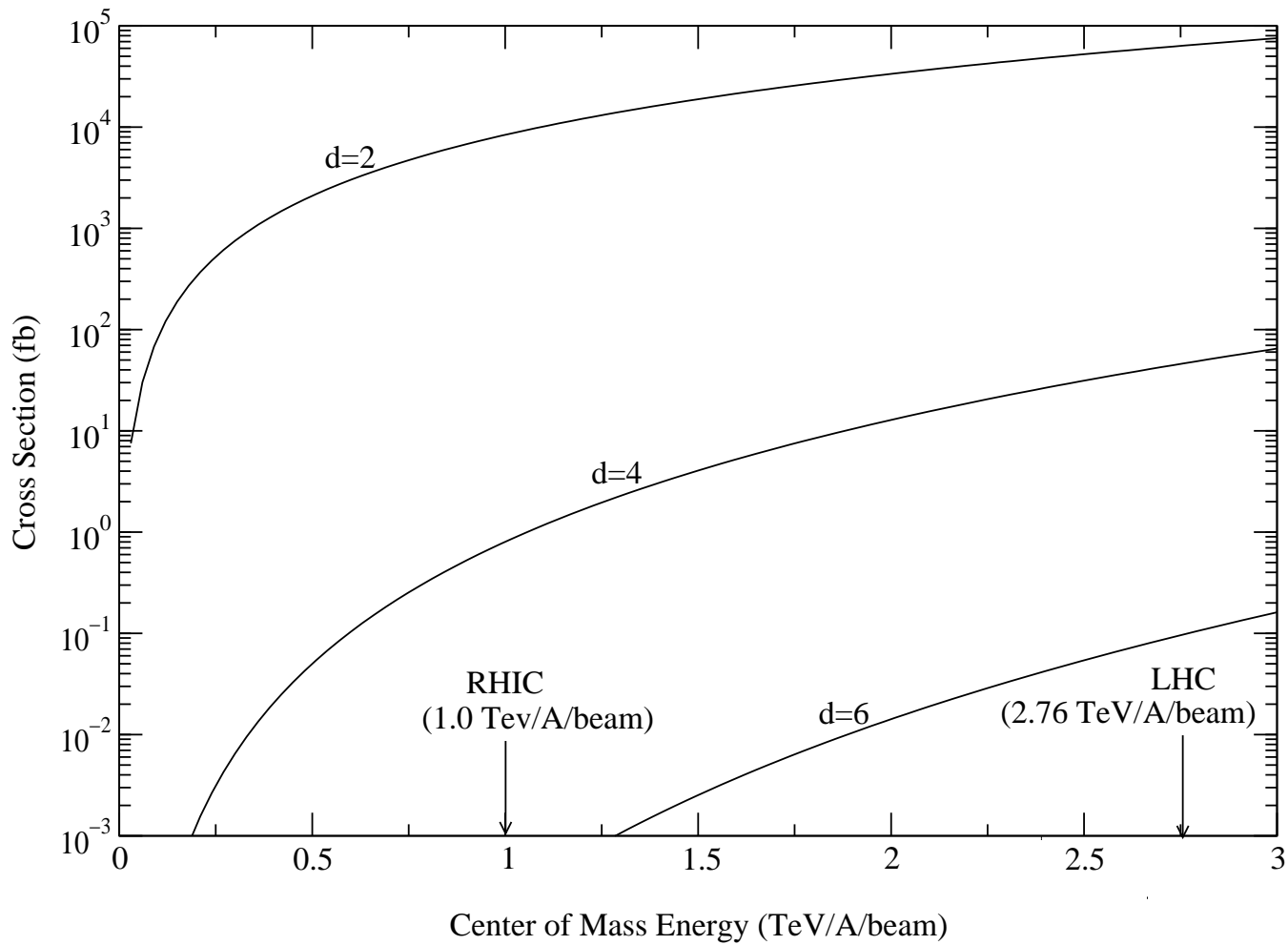}
\end{center}
\renewcommand{\figurename}{FIG.}
\caption{\label{cacaG} Cross section for graviton production in Ca-Ca collisions ($CaCa \to CaCa \gamma \gamma \to CaCaG$) at three different values of $d$, the number of large compact dimensions.}
\end{figure}

The cross section curves are shown in Fig. \ref{pbpbG} (for $PbPb \to PbPbG$)
and Fig. \ref{cacaG} (for $CaCa \to CaCaG$) for values of $d$ = $2$, $4$ and
$6$, along with indications of the machine energies of the planned RHIC
(1 TeV/A/beam) and LHC (2.76 TeV/A/beam). There is clearly an enhancement of
several orders of magnitude compared to the cross section for the same process
via electron-positron scattering, at least for certain ranges of parameters.
However, it must be pointed out that the work presented in this paper neglected to
take into account various complicating factors that might potentially be of great significance.
The most obvious is mentioned in \cite{Atw} --- that we do not have a proper quantum theory of gravity to work with, so there are necessarily uncertainties in this regard, particularly in utilizing the density of modes function (Eq. (\ref{two})). 
Another one is that a definitive identification of the graviton signature is precluded by the oversimplified nature of the formulation we used.
We assumed the usual Weizs\"{a}cker-Williams scenario, wherein the scattering angles of the interacting particles are always negligibly small, which means that the nuclei contributing to graviton production cannot be distinguished from the other nuclei in the accelerator beams.
Since the graviton couples only very weakly to ordinary matter, the signature for the overall reaction would be missing mass-energy, and therefore a definitive experimental signature cannot be predicted.
This problem can be possibly remedied by relaxing the assumption that the photons are on-shell.
Presumably, though, the resulting calculations (which would involve such concepts as nuclear form factors, partons within quarks, and the hadronic structure of photons) would yield much smaller signal cross sections.
Furthermore, for nucleus-nucleus collisions, one must also take into account
limitations due to other effects such as luminosity and background. The $\gamma\gamma$ luminosity ${\mathcal L}$ in a heavy ion collider is generally suppressed by several orders of magnitude compared to that in an electron-positron collider;
compare ${\mathcal L} \sim 10^{32} \ cm^{-2}s^{-1}$ for $e^+e^-$ scattering at
LEP2 to ${\mathcal L} \sim5 \times 10^{26} \ cm^{-2}s^{-1}$ for Pb-Pb
collisions at LHC and ${\mathcal L} \sim5 \times 10^{30}\ cm^{-2}s^{-1}$
for Ca-Ca collisions at LHC \cite{Hen, Vid}. In addition, electron-positron
collisions are expected to be much cleaner experimentally compared to
nucleus-nucleus collisions because of the hadronic debris accompanying
processes of the latter type \cite{Vid}. Although, one could always trigger
against multiplicity to reject this background.
\medskip
\begin{center}
\Roman{chapter}. CONCLUSIONS
\end{center}
\medskip
In summary, we investigated graviton production via two different processes.
The first production mechanism was through $\gamma\gamma$
fusion in electron-positron scattering, and we summarized a calculation that
used the Feynman rules in the Weizs\"{a}cker-Williams leading log
approximation. The second process was graviton production via
$\gamma\gamma$ fusion in peripheral nucleus-nucleus collisions, where we
considered both $^{208}$Pb and $^{40}$Ca ions. We calculated the cross sections for
these reactions using the semiclassical Weizs\"{a}cker-Williams
method and found them to be comparable to, and in some cases substantially
greater than, that for the previous process.
But, we note that because of the oversimplified nature of our analysis, there are potentially great uncertainties in our results.
\medskip
\begin{center}
ACKNOWLEDGEMENTS
\end{center}
\medskip
This work was supported in part by a graduate school dissertation fellowship from UW-Milwaukee and by funding from the National Space Grant College and Fellowhsip Program through the Wisconsin Space Grant Consortium.

\newpage

\newpage

\newpage

\end{document}